\documentstyle[10pt]{article}

\def\drm{d}
\def\text#1{\mbox{#1}}
\def\Mp{m_{\text{pl}}}

\def\bfl{\begin{flushleft}}
\def\efl{\end{flushleft}}
\def\bfr{\begin{flushright}}
\def\efr{\end{flushright}}
\def\bc{\begin{center}}
\def\ec{\end{center}}
\def\bqt{\begin{quote}}
\def\eqt{\end{quote}}
\def\be{\begin{equation}}
\def\ee{\end{equation}}
\def\ba{\begin{eqnarray}}
\def\ea{\end{eqnarray}}
\def\nn{\nonumber }

\begin{document}

~\\
~\\
~\\
~\\
~\\
\bfr
gr-qc/9709030
\efr
\bc
{\LARGE \bf
Theory of thin shells in general relativity:\\
Equivalence of direct and Wheeler-DeWitt quantizations
}

~~\\
{\large Konstantin G. Zloshchastiev}\\
~~\\
Department of Theoretical Physics,
Dnepropetrovsk State University,\\
Nauchniy lane 13, Dnepropetrovsk 320625,
Ukraine.\\

\ec

~~\\

\abstract{
We justify the way of the direct quantization which means 
immediate quantization of a conservation law.
It is shown that this approach is equivalent to introducing the super 
Hamiltonian
on a minisuperspace in spirit of the Wheeler-DeWitt's approach.
Then we will have: 
all values are observable and have an obvious physical meaning and
well-defined application domain; 
wave function is well-defined without time slicing
and often can be exactly obtained; 
we can take off major mathematical troubles, and 
therefore, more complicated models
can be considered exactly without the perturbation theory.
}

~\\



~\\
~\\
~\\
~\\
~\\
~\\
~\\
~\\
~\\
~\\
~\\
~\\
~\\
~\\
~\\
~\\

Address for hard correspondence:\\
K. G. Zloshchastiev\\
Metrostroevskaya 5, fl.453\\
Dnepropetrovsk 320128\\
Ukraine\\

Email:\\
zloshch@hotmail.com\\
\large
\newpage


Beginning from the classical works \cite{dau,isr} 
the investigation of thin shells
in general relativity has got large development 
(see reviews \cite{sat}).
In present paper we will consider the general class of spherically 
symmetric shells 
with nonzero surface tension \cite{kuc} thereby main
attention will be paid to quantum aspects of the theory.

Let us consider a thin layer with surface stress-energy tensor of a
perfect fluid in general case (we use the units 
$\gamma=c=1$, where $\gamma$ is the gravitational constant)
\be
S_{ab}=\sigma u_a u_b + p (u_a u_b +~ ^{(3)}\!g_{ab}),
\ee
where $\sigma$ and $p$ are the surface energy density  and  pressure 
respectively, $u^a$ is the timelike unit vector, 
$^{(3)}g_{ab}$ is the metric on the shell.

We shall write the metrics of the spacetimes outside $\Sigma^{out}$ and 
inside $\Sigma^{in}$ the spherical shell  in the form
\be
ds_{{out}\choose{in}}^2 =
- \Phi^{\pm}(r) dt^2_\pm + \Phi^{\pm}(r)^{-1} dr^2 + r^2 d\Omega^2,
\label{eq2}
\ee
where $d\Omega^2$ is the metric of unit 2-sphere.
It is possible to show that if one uses the  proper time  $\tau$ of a shell,
then the energy conservation law  can be written as
\be
d~ \biggl(\sigma \sqrt{^{(3)}\!g}
   \biggr) =
-p~ d\biggl(\sqrt{^{(3)}\!g} \biggr) - \sqrt{^{(3)}\!g}~ 
\biggl[(T^{\tau n})^{out} - (T^{\tau n})^{in}\biggr] d\tau, 
\label{eq3}
\ee
where $T^{\tau n}=T^{\alpha\beta} u_\alpha n_\beta$ is the projection 
of stress-energy tensors in the $\Sigma^{out}$ and $\Sigma^{in}$ spacetimes
on the tangent and normal vectors,
$^{(3)}\!g=\det{(^{(3)}\!g_{ab})}$.
The worldsheet metric of a shell is 
\be
^{(3)}\!\drm s^2 = - \drm \tau^2 + R^2 \drm \Omega^2,          \label{eq3b}
\ee
where $R(\tau)$ turns to be the proper radius of the shell.

Imposing junction conditions across the shell,
we derive the equations of motion of such shells in the form 
\ba
&&\epsilon_+ \sqrt{\dot R^2+\Phi^+(R)} - 
\epsilon_- \sqrt{\dot R^2+\Phi^-(R)} = - \frac{m}{R}, \label{eq5}\\ 
&&m=4 \pi \sigma (R) R^2,                               \nn
\ea
where $\dot R=d R/d\tau$,
$m$ is the (effective) rest mass.
The choice of the pair $\{ \epsilon_+=\pm 1,~\epsilon_-=\pm 1 \}$ 
divides  all shells into the classes of black hole (BH) type  and traversable 
wormhole (WH) type shells.
Equations (\ref{eq3}) and (\ref{eq5}) together with the state 
equation  $p=p(\sigma,^{(3)}\!g)$ and choice of the signs $\epsilon_\pm$ 
uniquely determine the motion of the fluid shell.
For further we will assume $\sigma(R)$ as an already 
known function of the theory because in most cases
we can resolve the conservation
law (\ref{eq3}) independently of an equation of motion \cite{zlompla}.
Equation (\ref{eq5}) can also be rewritten without roots: double
squaring we obtain
\be
\dot R^2 = 
\left[
\frac{\Delta \Phi - m^2/R^2}
     {2 m/R}
\right]^2 - \Phi^-(R),             \label{eq6}
\ee
where $\Delta \Phi = \Phi^+(R) - \Phi^-(R)$.


At present there are many approaches to quantize thin 
shells that is connected with different 
way of constructing the Hamiltonian
structure \cite{not,hb,ber,hkk}.
Of course, (almost) all these methods give the different results 
(wave functions, spectra, etc), thus we can observe the non-equivalent 
theories in all their discouraging multiformity.
Therefore it is necessary to work out some unified approach by means of
which we could compare models.
Besides, much of the known approaches use the perturbation theory to 
obtain final results that can be dangerous within the framework of
highly non-linear general relativity.
Thus it would be very important if this unified approach
would be also maximum nonperturbative.

The pure minisuperspace approach which does not require any time 
slicing and
time gauge seems to be the most suitable 
candidate (see, e.g., Ref. \cite{vil}).
Indeed, 
if one takes a look at Eq.\ (\ref{eq6}), one can see no time as a
variable.
Moreover, the second order differential equations from which 
Eq.\ (\ref{eq6}) was 
obtained, also contain no time variable \cite{isr}.
Therefore, 
what is the reason for introducing a time (and related concepts) 
forcibly, all the more so it creates additional troubles?
Let us consider the two approaches within the frameworks of 
minisuperspace method.\\

(a) {\it Approach with effective mass}\\
Let us consider the minisuperspace model initially 
described by the Lagrangian
\be                                                \label{eq7}
L = \frac{m \dot R^2}{2} 
- \frac{m}{2} 
\left\{
       \Phi^- - 
       \left[
             \frac{\Delta\Phi - m^2/R^2}{2m/R}
       \right]^2
\right\},
\ee
where we 
mean the integral as a primitive, $m$ is the above-mentioned 
effective rest mass, $m=m(R)$.
The equation of motion  is thus
\[
\frac{\drm }{\drm \tau} (m \dot R) = 
\frac{m_{, R} \dot R^2}{2}
- 
\frac{1}{2}
\left\{
       m \Phi^- - 
       m
       \left[
             \frac{\Delta\Phi - m^2/R^2}{2m/R}
       \right]^2
\right\}_{,R},
\]
where ``$,R$'' means the derivative with respect to $R$.

Using time symmetry we can easy decrease an order of this 
differential equation and 
obtain 
Eq.\ (\ref{eq6}) up to additive constant which can be calibrated 
to zero (note, it is zero only on trajectories thus it is 
a constraint).
Therefore our Lagrangian indeed describes dynamics of thin shells.
The moment conjugate to the variable $R$ is 
$                                                
\Pi = m \dot R,
$
and the (super)Hamiltonian
is
\be                                                \label{eq8}
H = \frac{\Pi^2}{2 m} + 
\frac{m}{2} 
\left\{
       \Phi^- - 
       \left[
             \frac{\Delta\Phi - m^2/R^2}{2m/R}
       \right]^2
\right\}.
\ee
The prefix ``super'' 
means that, strictly speaking, $H$ is the functional 
defined 
on the superspace which is the space of all worldsheet 3-metrics 
(\ref{eq3b}) and matter field configurations acting on a shell.
Only 
due to spherical symmetry and presence of a single non-propagating  
degree of freedom we can obtain it as a standard Hamiltonian.

Recalling the above-mentioned zero constant  we obtain that
$H=0$ on the trajectories (\ref{eq6}).
Thus we mean this (super)Hamiltonian as a constraint, i.e.
\be                                            \label{eq9}
H \approx 0,
\ee
or, in the quantum case ($\Pi = - i \partial_R$) 
\be           \label{eq10}
H \Psi \approx 0,
\ee
and can directly quantize Eq.\ (\ref{eq6}) without any assumption 
about a time as was done
in the special case 
$m=\text{const}$ (dust shell, $\sigma = m/ 4 \pi R^2$) 
in Ref. \cite{zlo}.
Therefore, one always can  quantize Eq.\ (\ref{eq6}) directly 
without redundant motivations about a time gauge etc., 
just replace $\dot R^2$ by $\Pi^2/m^2$.

Then in Planckian units we obtain 
the wave equation for the Wheeler-DeWitt wave function $\Psi(R)$:
\be                                                \label{eq13analog}
\Psi^{\prime\prime} + 
m^2
\left\{
       \left[
             \frac{\Delta\Phi - m^2/R^2}{2m/R}
       \right]^2 -
       \Phi^- 
\right\}\Psi = 0,
\ee
from 
which (for bound states if they exist) we can obtain spectra for  
necessary values, e.g., total mass-energy $M_+$ \cite{jr} etc.

However, it should be noted that there exists another Lagrangian.\\

(b) {\it Approach with Planckian mass}\\
Let us consider the minisuperspace model described by the Lagrangian
\be                                                \label{eq11}
L = \frac{\Mp \dot R^2}{2} 
- \frac{\Mp}{2} 
\left\{
       \Phi^- - 
       \left[
             \frac{\Delta\Phi - m^2/R^2}{2m/R}
       \right]^2
\right\}.
\ee
In the same 
manner as was pointed out above one can see that the appropriate
equation of motion yields Eq. (\ref{eq6}).
The moment conjugate to the variable $R$ is $\Pi = \Mp \dot R$ 
and the (super)Hamiltonian is
\be                                                \label{eq12}
H = \frac{\Pi^2}{2 \Mp} + 
\frac{\Mp}{2} 
\left\{
       \Phi^- - 
       \left[
             \frac{\Delta\Phi - m^2/R^2}{2m/R}
       \right]^2
\right\}.
\ee
It is easy to see that Eqs.  (\ref{eq9}), (\ref{eq10}) are valid.
Therefore, one can quantize Eq.\ (\ref{eq6}) directly 
by means of changing $\dot R^2$ by $\Pi^2/\Mp^2$.
In Planckian units we therefore obtain the wave equation for the
Wheeler-DeWitt wave function $\Psi(R)$:
\be                                                \label{eq13}
\Psi^{\prime\prime} + 
\left\{
       \left[
             \frac{\Delta\Phi - m^2/R^2}{2m/R}
       \right]^2 -
       \Phi^- 
\right\}\Psi = 0.
\ee
Comparing 
Eqs. (\ref{eq13analog}) and (\ref{eq13}), we see that they are
not the same and lead to different results in general case;
however, we can not give absolute preference to any of them.
Moreover, our Lagrangians are  defined always up to some arbitrary
multiplicative function of $R$ which does not affect on the equation 
of motion (\ref{eq6}) but necessarily appears in Hamiltonians.
The arbitrariness of this function is nothing but the arbitrariness of 
the choice of an appropriate gauge.

Finally it should be pointed out how the wormhole/blackhole topology 
should be taken into account at quantization.
Indeed from Eqs. (\ref{eq6}), (\ref{eq8}), (\ref{eq12}) 
it is evident that by 
double squaring we annihilated
the root signs $\epsilon$ which determine topology.
However at quantization 
we always can take into account topology because 
Eqs. (\ref{eq13analog}), (\ref{eq13}) should be supplemented by 
boundary conditions 
(e.g., at zero and spatial infinity) which are evidently 
determined by a
specific (wormhole or black hole) topology. 

Thus in present paper we worked out the minisuperspace
approach and performed nonperturbative canonical 
quantization of spherically 
symmetric singular hypersurfaces in general relativity.

\def\CMPh{Commun. Math. Phys.}
\def\EPL {Europhys. Lett.}
\def\JPh {J. Phys.}
\def\CJP {Czech. J. Phys.}
\def\LMPh{Lett. Math. Phys.}
\def\NPh {Nucl. Phys.}
\def\PhE {Phys.Essays}
\def\PhL {Phys. Lett.}
\def\PhR {Phys. Rev.}
\def\PhRL{Phys. Rev. Lett.}
\def\PhRp{Phys. Rep.}
\def\NCim{Nuovo Cimento}
\def\NuPB{Nucl. Phys.}
\def\GRG {Gen. Relativ. Gravit.}
\def\CQG {Class. Quantum Grav.}
\def\prp {report}
\def\Prp {Report}

\def\jn#1#2#3#4#5{{#1}{#2} {#3} {(#5)} {#4}}
\def\boo#1#2#3#4#5{{\it #1} ({#2}, {#3}, {#4}){#5}}
\def\prpr#1#2#3#4#5{{``#1,''} {#2}{#3}{#4}, {#5} (unpublished)}

\end{document}